\documentclass[conference]{IEEEtran}
\IEEEoverridecommandlockouts
\pdfoutput=1
\usepackage{cite}
\usepackage{amsmath,amssymb,amsfonts}
\usepackage{algorithmic}
\usepackage{graphicx}
\usepackage{textcomp}
\usepackage{xcolor}
\def\BibTeX{{\rm B\kern-.05em{\sc i\kern-.025em b}\kern-.08em
    T\kern-.1667em\lower.7ex\hbox{E}\kern-.125emX}}

\usepackage{color,soul}

\usepackage{textcomp}
\allowdisplaybreaks

\begin{document}

\title{A Game Theoretic Approach for Demand Response Allocation among Strategic Prosumers in Regulated Distribution Utilities}

\author{\IEEEauthorblockN{Sayyid Mohssen Sajjadi}
\IEEEauthorblockA{\textit{School of Electrical, Computer and Energy Engineering} \\
\textit{Arizona State University}\\
Tempe, USA \\
}
\and
\IEEEauthorblockN{Meng Wu}
\IEEEauthorblockA{\textit{School of Electrical, Computer and Energy Engineering} \\
\textit{Arizona State University}\\
Tempe, USA \\
}

}

\maketitle

\begin{abstract}
This paper studies an optimal allocation of demand response (DR) provisions among strategic photovoltaic (PV) prosumers, through third-party DR providers which operate within the territory of the regulated distribution utility. A game theoretic model, consisting of a sequential game coupled with a simultaneous game, is proposed to capture the competition and interaction among PV prosumers, the third-party DR providers, and the utility company. An iterative approach is proposed to solve for the Nash equilibrium of the game.	With this game theoretic model, the DR provision quantities for strategic PV prosumers are optimally determined, considering the profit maximizations of the PV prosumers, the DR providers, and the utility company. Case studies on an IEEE test system verifies the proposed model and the solution approach.

\end{abstract}


\section{Introduction}
The penetration of distributed energy resources (DERs) is increasing significantly in both deregulated energy markets and regulated utilities, which offers great opportunities for third-party demand response (DR) providers to aggregate the flexibility of individual prosumers with DERs and provide grid services through regulated and deregulated markets \cite{b1,b2}. To optimally incentivize DR provisions from individual prosumers and fully utilize the capability of DR providers for grid services provision, it is critical to understand the interactions among the strategic behaviors of the individual prosumers, the third-party DR providers, and the operations of the regulated utilities and deregulated markets. \cite{b3}

Several existing works consider the above interactions in their studies.
A price-based DR is proposed in \cite{b4}, \cite{b5}. In \cite{b4}, the price-based DR is used to adjust supply and demand imbalance in microgrids, where microturbines are utilized in the DR program and their outputs are modified in hourly basis. In \cite{b5}, PV prosumers are integrated into the microgrid where the PV price is determined based on the supply and demand ratio (SDR). The end-users willingness is also taken into consideration in \cite{b5}, in which designing the inconvenience cost for participating in the DR program is desired. The end-users willingness to participate in the DR program is also studied in \cite{b6,b7}. A list of price plans is provided in \cite{b6} to encourage consumers to participate in the DR program. Reference \cite{b7} aims to minimize the inconvenience cost of consumers without providing incentives to DR participants. Instead, a wide range of household appliances are considered in the DR program in \cite{b7}, which allows consumers to choose their consumption patterns based on their willingness. In \cite{b17}, a game theoretic approach is proposed to capture the interactions among wholesale market participants with DR resources. In \cite{b18}, a Stackelberg game model is proposed to study the hierarchical decision making problem in smart grids.

This paper focuses on the problem of optimally allocating DR provisions among strategic PV prosumers controlled by third-party DR providers within the territory of the regulated distribution utility company. To study the strategic profit-seeking behavior of different PV prosumers, the DR providers, and the utility company, a game theoretic model is proposed consisting of a sequential (Stackelberg) game and a simultaneous game coupled together. The sequential game models the interaction between the utility and the DR provider, and the simultaneous game models the interaction between strategic PV prosumers under the same DR provider. The PV prosumers participate in third-party DR programs and sell their surplus PV generation as well as DR quantity to the DR providers. The DR providers then sell the aggregated DR quantity (bought from the PV prosumers) to the utility. Dynamic pricing models are proposed for the utility to incentivize aggregated DR provision from the DR provider, and for the DR provider to incentivize the DR participation of PV prosumers. The utility reduces its operation cost by achieving load reductions during DR events. An iterative approach is proposed to solve for the Nash equilibrium of the coupled games, at which the prosumers' and DR providers' benefits are maximized while the utility's operation cost is reduced considering all the interactions among different entities. To the best of our knowledge, the proposed coupled game structure and studies on the interactions/competitions among different strategic PV prosumers, the third-party DR providers, and the utility company have not been addressed in the existing literature.

The rest of this paper is organized as follows. Section II presents the structure of the coupled games. Section III models different entities in the games. Section IV presents the iterative approach for solving the coupled games. Section V presents the case study results. Section VI concludes this paper.

\begin{figure}
	\centering
	\includegraphics[width=0.45\textwidth]{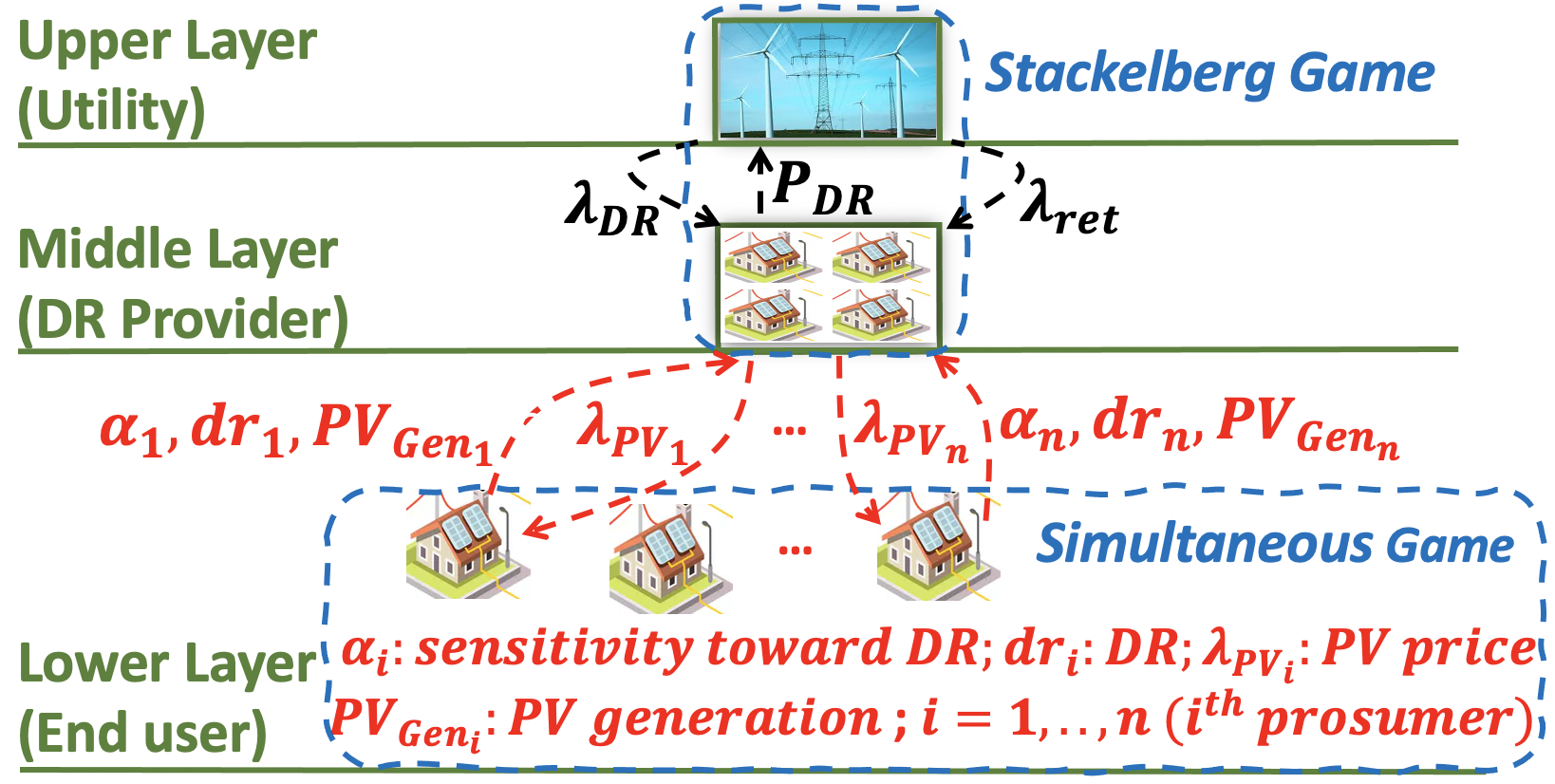}
	\caption{Structure of the proposed coupled game model.}
	\label{Market Structure}
	\vspace{-0.5cm}
\end{figure}

\section{Structure of the Coupled Game Model}
The structure of the proposed coupled game model is shown in Fig.~\ref{Market Structure} with three coupled layers - the utility as the upper layer, the DR providers as the middle layer, and the PV prosumers (end users) as the lower layer. Each DR provider has $n$ PV prosumers selling both their surplus PV generation and DR quantity. The DR provider is paid by the utility at the DR price $\lambda_{{DR}_i}$ for the DR quantity provided by $i^{th}$ PV prosumer within its footprint. The DR provider then pays $i^{th}$ PV prosumer at the PV price $\lambda_{{PV}_i}$. The utility charges the inelastic loads not participating in any DR program at the time-of-use retail rate $\lambda_{ret}$. The price signals in this model assume there is no DR provision before the DR event scheduled by the utility. The DR provider also receives the sensitivity value $\alpha_i \in [0,1]$ from $i^{th}$ PV prosumer, which represents the sensitivity/willingness of the PV prosumer to provide DR services. Each PV prosumer also sends its PV generation quantity, PV generation cost, and maximum DR capability to the DR provider.

The utility and the DR provider are coupled together through the DR prices $\lambda_{{DR}_i}$ and compete with each other through the sequential (Stackelberg) game, for the optimal amount of aggregated DR quantity which satisfy the utility's cost reduction objective and the DR provider's profit maximization objective. The DR provider and the end users are coupled together through the PV prices $\lambda_{{PV}_i}$.  The DR provider runs a simultaneous game among the PV prosumers, which simulate the prosumers' strategic competition behaviors with each other, after receiving the PV price signals. This simultaneous game determines the optimal DR quantity for each PV prosumer, considering the profit maximization objectives of the DR provider and the PV prosumers, as well as the competitions among all the PV prosumers. Through these coupled interactions, the utility, the DR provider, and the PV prosumers jointly determine the optimal allocations of the total DR quantity among all the PV prosumers, considering the cost reduction/profit maximization objectives of all the entities.

\section{Problem Formulation}
\subsection{Lower Layer Mathematical Modeling}
The lower layer models the PV prosumer whose utility function contains two components, the inconvenience cost and the profit of selling surplus PV power to the PV provider in the middle layer.

\subsubsection{The Inconvenience Cost}
This component represents prosumer's willingness to curtail or shift load during the DR event. This paper adopts the equivalent cost of inconvenience as a function of DR quantity, which is formulated in \cite{b5}.
\begin{equation}
Inc_i^t=\alpha_i\cdot (p_i^t-x_i^t)^2
\label{original invonvenience cost}
\end{equation}
subject to:
\begin{equation}
    dr_i^t = p_i^t - x_i^t
\label{eqn:DR}
\end{equation}
\begin{equation}
0\leq x_i^t \leq p_i^t
\label{eqn:DR constraint}
\end{equation}
where $\alpha_i$, $p_i^t$, $x_i^t$, and $dr_i^t$ denote the prosumer's sensitivity/willingness for DR provision, the desired power consumption during the DR event (before providing any DR quantity), the adjusted power consumption during the DR event (after providing the DR quantity), and the DR quantity provided by $i^{th}$ prosumer at time $t$, respectively. To capture the interaction/competition between different prosumers, we assign a per unit parameter, $\theta_i^t$, to each prosumer, and re-formulate the inconvenience cost in \cite{b5} as follows.
\begin{equation}
\theta_i^t=\frac{\frac{1}{(p_i^t-x_i^t)}}{\sum \limits_{k=1}^n\frac{1}{ (p_k^t-x_k^t)}}
\label{eqn:theta 1}
\end{equation}
where $n$ is the number of prosumers within the DR provider's footprint. From (\ref{original invonvenience cost}) and (\ref{eqn:theta 1}), it is clear that $\theta_i^t$ is proportional to $\frac{1}{(p_i^t-x_i^t)}$ and $Inc_i^t$ is proportional to $(p_i^t-x_i^t)$, indicating $Inc_i^t$ being proportional to $\frac{1}{\theta_i^t}$. Therefore, the inconvenience cost in \cite{b5} is re-formulated as follows.
\begin{equation}
\begin{aligned}
Inc_i^t &= \alpha_i \cdot (p_i^t-x_i^T)^2 \cdot \frac{1}{\theta_i} =\alpha_i\cdot \frac{(p_i^t-x_i^t)^2}{\frac{\frac{1}{ (p_i^t-x_i^t)}} {\sum \limits_{k=1}^n\frac{1}{(p_k^t-x_k^t)}}} \\
&=\alpha_i\cdot(p_i^t-x_i^t)^3 \cdot \left( \sum \limits_{k=1}^n\frac{1}{(p_k^t-x_k^t) } \right)
\end{aligned}
\label{eqn:inc_new}
\end{equation}

In (\ref{eqn:inc_new}), the parameter $\theta_i^t$ is introduced to model the fact that the optimal decision taken by prosumer $i$ depends upon the decisions/strategies of other prosumers, which represents the potential competition among prosumers when they decide their optimal DR provisions simultaneously (through the simultaneous game). When there is no game considered among prosumers, i.e., the inconvenience cost in (\ref{original invonvenience cost}) is used, the DR provider would prefer purchasing DR quantity from the prosumers to maximize its own utility function (when the DR participation is the highest) and the prosumers passively accept the DR provider's decision. When the simultaneous game is considered between prosumers, i.e., the inconvenience cost in (\ref{eqn:inc_new}) is used, each prosumer adjusts its DR provision strategies to maximize its own utility function (i.e., minimizing its net cost), such that it could be selected by the DR provider to sell its DR quantity during the DR event and be paid at the PV price. Upon convergence of the game, a Nash equilibrium will be reached at which all the prosumers adjust their strategies to the point that no single prosumer could improve its own benefit without decreasing the other prosumers' benefits.

\subsubsection{The Profit of Selling Surplus PV Power to The DR Provider}
This component can be modeled as follows.

\begin{equation}
    Profit_{{PV}i}^t = \lambda_{{PV}_i}^t \cdot (PV_{{Gen}_i}^t-x_i^t)
\label{eqn:pv_profit}
\end{equation}
where $Profit_{{PV}i}^t$, $\lambda_{{PV}_i}^t$, and $PV_{{Gen}_i}^{t}$ denote the profit of selling surplus PV power, the PV price paid by the DR provider, and the PV power generation for $i^{th}$ prosumer at time $t$, respectively.

In (\ref{eqn:pv_profit}), the dynamic PV price, $\lambda_{{PV}_i}^t$, is modeled as a function of the supply and demand ratio (SDR) \cite{b5}. The SDR for $i^{th}$ prosumer at time $t$, $SDR_i^t$, can be written as follows.
\begin{equation}
    SDR_i^t = \frac{PV_{{Gen}_i}^t}{x_i^t}
\label{SDR}    
\end{equation}

The dynamic PV price function, $\lambda_{{PV}_i}^t(SDR_i^t)$, is designed based on the following two principles: 1) the PV price is bounded between the PV generation cost and the utility's time-of-use retail rate, i.e., $(PV_{GC}^t<\lambda_{PV}^t\leq \lambda_{ret}^t)$, where $PV_{GC}^t$ denotes the PV generation cost; 2) the relation between the PV price and the SDR is inverse-proportional \cite{b5}. To guarantee the PV prosumers' profit, their PV prices need to be maintained greater than their PV generation costs. The aforementioned principles can be 
formulated as follows. Note that the subscript $i$ is omitted below for the purpose of simplicity.
\begin{equation}
\begin{cases}
    PV_{GC}^t<\lambda_{PV}^t\leq \lambda_{ret}^t  \\
    \lambda_{PV}^t = 0 & (\text{for~} SDR^t \leq1) \\
    \lim_{SDR^t\to1^+}{\lambda_{PV}^t = \lambda_{ret}^t} \\
    \lim_{SDR^t\to\infty}{\lambda_{PV}^t = PV_{GC}^t} & (\text{for~} SDR^t \gg1)
\end{cases}
\label{eq:pv_price_principle}
\end{equation}

Based on the principles in \eqref{eq:pv_price_principle}, the fitting function below is proposed which relates the PV price and the SDR.
\begin{equation}
\label{original function PV}
\lambda_{PV}^t(SDR^t) = \frac{a\cdot SDR^t}{b\cdot SDR^t+c}
\end{equation}

According to \eqref{eq:pv_price_principle}, we have:
\begin{equation}
\begin{cases}
\lim_{SDR^t\to1^+}{\lambda_{PV}^t(SDR^t)} =  \frac{a}{b+c}=\lambda_{ret}^t \\
\lim_{SDR^t\to\infty}{\lambda_{PV}^t(SDR^t)}= \frac{a}{b}=PV_{GC}^t
\end{cases}
\label{eq_10}
\end{equation}

Equation \eqref{eq_10} implies:
\begin{equation}
\begin{cases}
c=\frac{a}{\lambda_{ret}^t}-b \\
a=b\cdot PV_{GC}^t 
\end{cases}
\label{eq_11}
\end{equation}

Integrating \eqref{eq_11} into \eqref{original function PV}, we obtain:
\begin{align}
    \lambda_{PV}^t(SDR^t) &= \frac{b\cdot PV_{GC}^t\cdot SDR^t}{b\cdot SDR^t+\frac{a}{\lambda_{ret}^t}-b} \nonumber \\
    &= \frac{b\cdot PV_{GC}^t \cdot SDR^t}{b\cdot SDR^t+\frac{b \cdot PV_{GC}^t}{\lambda_{ret}^t}-b}  \label{eq_12}\\
    &= \frac{PV_{GC}^t \cdot SDR^t}{SDR^t+\frac{PV_{GC}^t}{\lambda_{ret}^t}-1} \nonumber\\
    &= \frac{\lambda_{ret}^t \cdot PV_{GC}^t \cdot SDR^t} {\lambda_{ret}^t \cdot SDR^t + (PV_{GC}^t - \lambda_{ret}^t)} \nonumber
\end{align}

Therefore, integrating other principles in \eqref{eq:pv_price_principle} into \eqref{eq_12}, the dynamic PV price from the DR provider to the PV prosumer is modeled as follows:
\begin{equation}
    \lambda_{{PV}_i}^t =
    \begin{cases}
    \frac{\lambda_{ret}^t\cdot PV_{{GC}_i}^t\cdot SDR_i^t}{\lambda_{ret}^t\cdot SDR_i^t + (PV_{{GC}_i}^t-\lambda_{ret}^t)}, \text{if $SDR_i^t>1$}
    \\
    0,\text{if $SDR_i^t\leq1$}
    \end{cases}
\label{lambda PV}
\end{equation}

In \eqref{lambda PV}, the DR provider calculates PV price when the prosumers' supply (surplus PV generation plus DR) is greater than their demand ($x_i^t$), and the PV price value is calculated based on the received DR provision signals ($x_i^t = p_i^t - dr_i^t$).

\subsubsection{Optimal Decision of Each PV Prosumer}
Each PV prosumer behaves strategically to minimize its net cost for DR provision:
\begin{align}
   \min_{x_i^t} \quad & \sum_{t=1}^{24} (Inc_i^t - Profit_{{PV}i}^t) \label{eqn:utility function of end-user}\\
   \textrm{s.t.} \quad & \textrm{\eqref{eqn:DR constraint}} \nonumber
\label{OF_End-user}
\end{align}

The above minimization problem is a convex optimization problem and can be solved using general convex optimization solvers. Proof of convexity is omitted due to space limitation.

\subsection{Middle Layer Mathematical Modeling}
The DR provider's objective is to maximize its net profit. Its profit depends on the total DR quantity received from the PV prosumers in the lower layer. Therefore, the best strategy for the DR provider would be to encourage prosumers to participate in the DR program. The encouragement in this paper takes place through the dynamic PV price modeling mechanism proposed in previous section. The DR provider's profit at time $t$, $Profit^t$, can be modeled as below.
\begin{equation}
    Profit_{DR}^t =  \sum \limits_{i=1}^{n}(\lambda_{{DR}_i}^t - \lambda_{{PV}_i}^t) \times (PV_{{Gen}_i}^t - x_i^t)
    \label{eqn_14}
\end{equation}

In this paper, the dynamic DR price paid to the DR provider from the utility is modeled using mechanisms similar to \eqref{eq:pv_price_principle}-\eqref{lambda PV}. The following design principles are satisfied.

\begin{equation}
\begin{cases}
\lambda_{PV}^t<\lambda_{DR}^t\leq \lambda_{ret}^t \\
\lambda_{DR}^t = 0 & \text{for~} SDR^t \leq1 \\
\lim_{SDR^t\to1^+}{\lambda_{PV}^t = \lambda_{ret}^t} \\
\lim_{SDR^t\to\infty}{\lambda_{DR}^t = \lambda_{PV}^t} & \text{for~} SDR^t \gg1
\end{cases}
\end{equation}

Consequently, the dynamic DR price is modeled as follows.
\begin{equation}
    \lambda_{{DR}_i}^t =
    \begin{cases}
    \frac{\lambda_{ret}^t\cdot \lambda_{{PV}_i}^t\cdot SDR_i^t}{\lambda_{ret}^t\cdot SDR_i^t + (\lambda_{{PV}_i}^t-\lambda_{ret}^t)}, \text{if $SDR_i^t>1$}
    \\
    0,\text{if $SDR_i^t\leq1$}
    \end{cases}
    \label{eqn_17}
\end{equation}
where $\lambda_{{DR}_i}^t$ is the price utility pays the DR provider at time $t$ for providing DR from prosumer $i$; $\lambda_{{PV}_i}^t$ is the PV price DR provider pays the prosumer $i$ at time $t$ for its DR provision.

In \eqref{eqn_17}, the lowest DR price is $\lambda_{{PV}_i}^t$, and it happens when the $SDR_i^t$ goes toward infinity. In practice, the $SDR_i^t$ is a limited value, given the fact that the DR capability of a prosumer is limited. This guarantees the DR provider gets paid by the utility at a price higher than the price it pays the prosumers for their surplus PV and DR quantities.

\subsection{Upper Layer Mathematical Modeling}
The utility's objective is to maximize the system operating cost reduction through DR services. This paper adopts the below quadratic cost function in \cite{b8,b9} to model the utility's operating cost.
\begin{equation}
    Cost^t(P_{system}^t) = c_0 + c_1\cdot P_{system}^t + c_2 \cdot (P_{system}^t)^2
\end{equation}
where $P_{system}^t$ is the desired power consumption (before DR events) across the distribution system at time $t$; $c_0$, $c_1$ and $c_2$ are the coefficients of the cost function. After having a $DR$ event, the cost function could be written as follows:
\begin{equation}
Cost^t(Q^t)=c_0 + c_1\cdot Q^t+c_2 \cdot (Q^t)^2
\end{equation}
where $Q^t = P_{system}^t-\sum \limits_{i=1}^{n} PV_{Gen}^t-\sum \limits_{i=1}^{n} dr_i^t$ denotes the total adjusted load across the distribution system at time $t$, after the DR event. The utility's profit from the DR event, $Profit_{UC}^t$, can be developed as follows. 
\begin{equation}
Profit_{UC}^t=Cost^t(P_{system}^t) - Cost^t(Q^t)
\end{equation}

\section{The Iterative Solution Procedure}
The proposed iterative solution procedure is as follows.

\textbf{Step 1}: Share the initial data (e.g. $PV_{{Gen}_i}^t$, $PV_{{GC}_i}^t$, $\lambda_{ret}^t$, and $p_i^t$) among game players. Note that $x_i^t = p_i^t$ and $dr_i^t = 0$ at the first iteration.

\textbf{Step 2}: Calculate $SDR_i^t$.

\textbf{Step 3}: The DR provider provides the PV prosumers with $\lambda_{{PV}_i^t}$ (see \eqref{lambda PV}) and share this value with the utility. The utility then provides the DR-provider with $\lambda_{{DR}_i^t}$ (see \eqref{eqn_17}).

\textbf{Step 4}: The PV prosumers update their optimal DR provisions through the simultaneous game.

\textbf{Step 5}: The utility and the DR provider update their objective functions. 
	
\textbf{Step 6}: If the convergence criteria in (\ref{eqn_21}) are satisfied, the iteration stops. Otherwise, go to Step 2.
\begin{equation}
\begin{cases}
    |dr_{i_{new}}^t - dr_{i_{old}}^t| \leq \epsilon_1 \\
    |Profit_{DR_{new}}^t - Profit_{DR_{old}}^t| \leq \epsilon_2 \\
    |Profit_{UC_{new}}^t - Profit_{UC_{old}}^t| \leq \epsilon_3
\end{cases}
\label{eqn_21}
\end{equation}
where $\epsilon_1$, $\epsilon_2$, and $\epsilon_3$ denote small positive thresholds for the criteria; the subscripts $old$ and $new$ denote the values obtained from the previous and current iterations, respectively.

Our proposed model follows AC saver thermostat program at San Diego Gas \& Electric Company (SDG\&E) \cite{b10} where AC thermostat will be adjusted remotely during DR event from noon to 9:00pm, for no more than four hours.

\section{Simulation results and discussion}
\begin{figure}[]
\centering
\includegraphics[width=0.4\textwidth, height = 3.1cm]{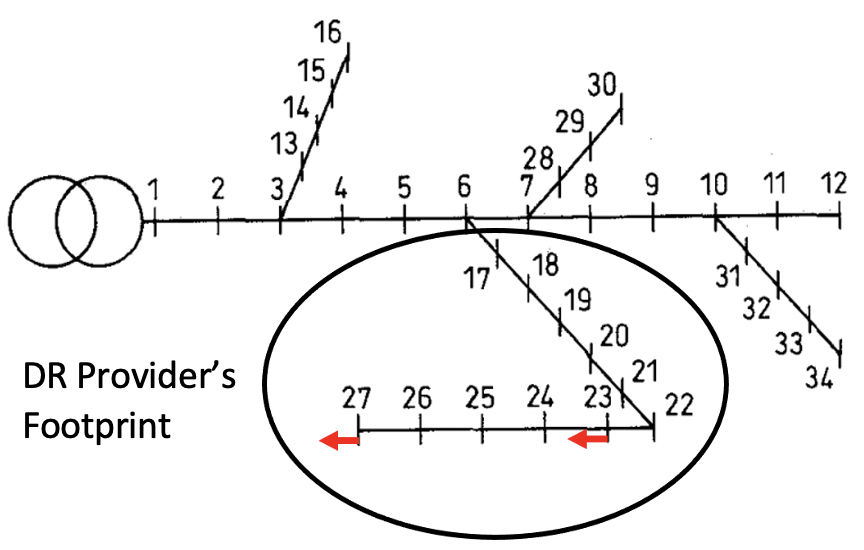}
\caption{IEEE 34 bus test system}
\label{Case study}
\end{figure}

\begin{figure}[]
\centerline{\includegraphics[width=0.4\textwidth, height = 3.3cm]{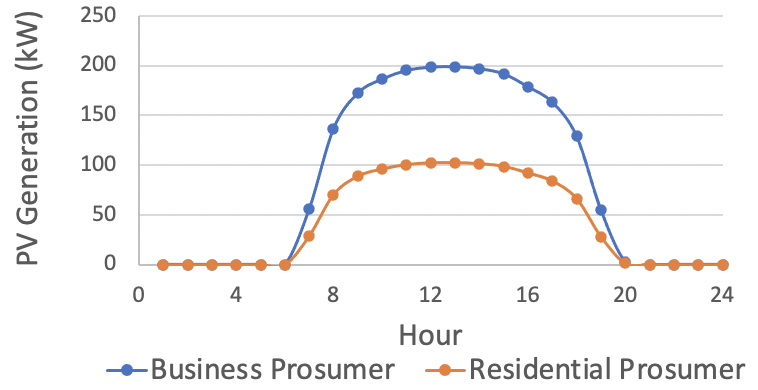}}
\caption{PV generation profile for residential and business prosumers}
\label{PV data}
\end{figure}

The proposed model is tested on the IEEE 34-Bus test system \cite{b11} shown in Fig.~\ref{Case study}. The DR provider's footprint includes bus 23 and bus 27 where bus 23 and bus 27 are assumed to be the business prosumer and the residential prosumer, respectively. The load data is shifted such that it follows the pattern with the peak hours from 4:00pm to 9:00pm, off-peak hours from 6:00am to 3:00pm and from 10:00pm to 12:00am, and super off-peak hours from 01:00am to 05:00am, based on the price program at SDG\&E during summer time. The residential and business prosumers are assumed to have 17 residential buildings with a 6.2 kW PV system each, and one business building with a 200 kW PV system, respectively, which is consistent with \cite{b12}. The PV generation data is adopted from California ISO on July 14, 2020, and is scaled as shown in Fig.~\ref{PV data}. The time-of-use retail rate for residential consumers is adopted from SDG\&E. The retail rate for business consumers is computed based on the ratio between the retail rates for residential and business consumers in Salt River Project (SRP) \cite{b14,b15}, as the SDG\&E's retail rate for business consumers is not public online. The PV generation costs are adopted from \cite{b16}. The Maximum DR capability for prosumers is assumed to be 10\% of the peak load. The utility's cost function coefficients $c_0$, $c_1$, and $c_2$ are 4207.5, -6.74, 0.0029, and are computed based on three load values and their corresponding prices \cite{b8}, \cite{b9}. The $\alpha$ values for residential and business prosumers is set to be 0.8 and 0.9, respectively, ensuring that the prosumers' DR range is within $0-10\%$ of their peak load.

\begin{figure}[t]
\centerline{\includegraphics[width=0.4\textwidth, height = 3.7cm]{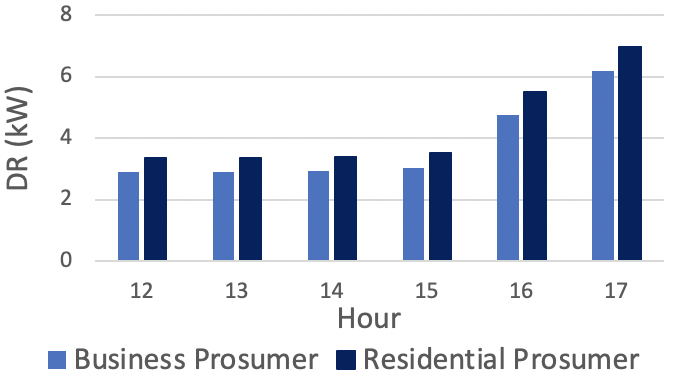}}
\caption{Optimal DR quantity of residential and business prosumers}
\label{DR kW}
\end{figure}

\begin{figure}[t]
\centerline{\includegraphics[width=0.4\textwidth, height = 3.7cm]{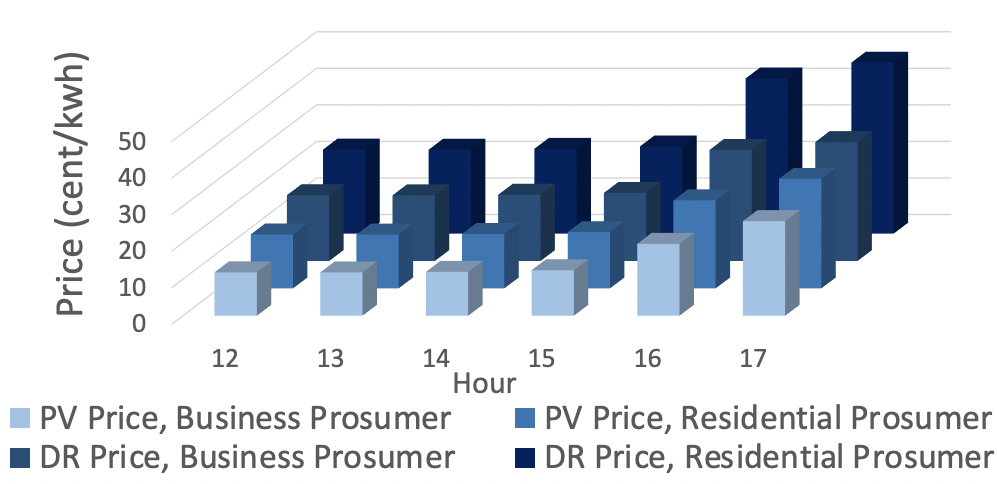}}
\caption{PV price and DR price during DR event}
\label{Price}
\end{figure}

\begin{figure}[t]
\centerline{\includegraphics[width=0.4\textwidth, height=3.5cm]{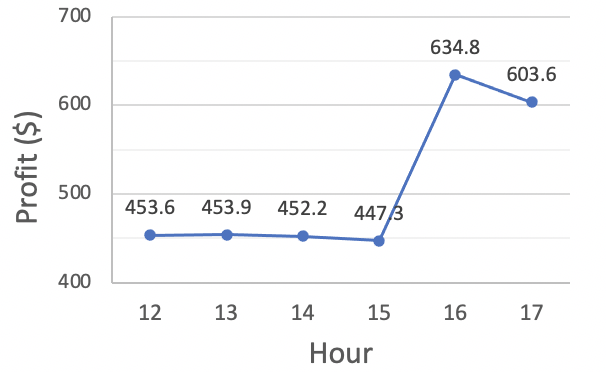}}
\caption{Utility profit during DR event}
\label{Utility profit}
\end{figure}

Fig.~\ref{DR kW} shows the optimal DR quantity of the residential and business prosumers. Fig.~\ref{Price} shows the PV price and DR price of the residential and business prosumers. It shows that the DR quantity increases as retail rates increase during peak hours, and the prosumers obtain the most profit by selling PV power and providing DR quantity during peak hours when the PV prices are high. The prosumer's behaviour during peak hours verifies the effectiveness of the proposed mechanism. The prosumers' participation in the DR program at 04:00pm and 05:00pm are the highest while the PV system generates less power at these hours compared to previous hours. This is because the PV prices at 04:00pm and 05:00pm are higher compared to previous hours, therefore the prosumers are strongly encouraged to participate in the DR program. The PV generation on and after 06:00pm is much less than the load. Therefore, there is no surplus PV power to be sold and the DR participation is zero. It is shown in Fig.~\ref{Price} that the DR provider gets paid at higher prices at 4:00pm and 5:00pm when the retail rate is the highest. Fig.~\ref{Utility profit} shows the utility's profit during the DR event, which provides insights to the utility on the appropriate hours to schedule DR events.

\section{Conclusion}
This paper proposed a three-layer game theoretic approach to study the optimal allocation of DR provisions among the PV prosumers through the third-party DR provider resided in the territory of the regulated distribution utility. Interactions/competitions among PV prosumers, the DR provider, and the utility company are modeled using the coupled sequential and simultaneous games. The dynamic PV price modeling and dynamic DR price modeling are formulated in a way that prosumers are encouraged to actively participate in the DR program. The effectiveness of the proposed work is verified through case studies on an IEEE test system.

Future work could focus on considering closer-to-practice models for different entities in the games, and developing better pricing mechanisms for the utility and the DR provider.



\begin{thebibliography}{00}
\bibitem{b1} J. Campos do Prado and W. Qiao, "A vision of the next-generation retail electricity market in the context of distributed energy resources," 2018 IEEE Power \& Energy Society Innovative Smart Grid Technologies Conference (ISGT), Washington, DC, 2018, pp. 1-5.

\bibitem{b2} P. Li, H. Wang and B. Zhang, "A Distributed Online Pricing Strategy for Demand Response Programs," in IEEE Transactions on Smart Grid, vol. 10, no. 1, pp. 350-360, Jan. 2019.

\bibitem{b3} P. Jacquot, O. Beaude, S. Gaubert and N. Oudjane, "Analysis and Implementation of an Hourly Billing Mechanism for Demand Response Management," in IEEE Transactions on Smart Grid, vol. 10, no. 4, pp. 4265-4278, July 2019.

\bibitem{b4} C. Zhang, Y. Xu, Z. Y. Dong and K. P. Wong, "Robust Coordination of Distributed Generation and Price-Based Demand Response in Microgrids," in IEEE Transactions on Smart Grid, vol. 9, no. 5, pp. 4236-4247, Sept. 2018.

\bibitem{b5} N. Liu, X. Yu, C. Wang, C. Li, L. Ma and J. Lei, "Energy-Sharing Model With Price-Based Demand Response for Microgrids of Peer-to-Peer Prosumers," in IEEE Transactions on Power Systems, vol. 32, no. 5, pp. 3569-3583, Sept. 2017.

\bibitem{b6} H. Yang, J. Zhang, J. Qiu, S. Zhang, M. Lai and Z. Y. Dong, "A Practical Pricing Approach to Smart Grid Demand Response Based on Load Classification," in IEEE Transactions on Smart Grid, vol. 9, no. 1, pp. 179-190, Jan. 2018.

\bibitem{b7} O. Alrumayh and K. Bhattacharya, "Flexibility of Residential Loads for Demand Response Provisions in Smart Grid," in IEEE Transactions on Smart Grid, vol. 10, no. 6, pp. 6284-6297, Nov. 2019.

\bibitem{b17} M. Yu, S. H. Hong, Y. Ding and X. Ye, "An Incentive-Based Demand Response (DR) Model Considering Composited DR Resources," in IEEE Transactions on Industrial Electronics, vol. 66, no. 2, pp. 1488-1498, Feb. 2019.

\bibitem{b18} W. Tushar et al., "Three-Party Energy Management With Distributed Energy Resources in Smart Grid," in IEEE Transactions on Industrial Electronics, vol. 62, no. 4, pp. 2487-2498, April 2015, doi: 10.1109/TIE.2014.2341556.

\bibitem{b8} P. Jacquot, O. Beaude, S. Gaubert and N. Oudjane, "Demand response in the smart grid: The impact of consumers temporal preferences," 2017 IEEE International Conference on Smart Grid Communications (SmartGridComm), Dresden, 2017, pp. 540-545.

\bibitem{b9} G. Tsaousoglou, K. Steriotis, N. Efthymiopoulos, P. Makris and E. Varvarigos, "Truthful, Practical and Privacy-Aware Demand Response in the Smart Grid via a Distributed and Optimal Mechanism," in IEEE Transactions on Smart Grid, vol. 11, pp. 3119-3130, July 2020.

\bibitem{b10} [Online]. Available:
https://www.sdge.com/residential/savings-center/energy-saving-programs/reduce-your-use/reduce-your-use-thermostat

\bibitem{b11} M. Chis, M. M. A. Salama and S. Jayaram, "Capacitor placement in distribution systems using heuristic search strategies," in IEE Proceedings - Generation, Transmission and Distribution, vol. 144, no. 3, pp. 225-230, May 1997.

\bibitem{b12} R. Fu, D. Feldman, and R. Margolis, "US solar photovoltaic system cost benchmark: Q1 2018," No. NREL/TP-6A20-72399. National Renewable Energy Lab.(NREL), Golden, CO (United States), Nov. 2018.

\bibitem{b14} [Online]. Available: https://www.srpnet.com/prices/business/tou.aspx

\bibitem{b15} [Online]. Available: https://www.srpnet.com/prices/home/tou.aspx


\bibitem{b16} [Online]. Available:
https://www.energy.gov/sites/prod/files/2016/12/f34 /SunShot

\end{thebibliography}
\end{document}